\documentclass[twocolumn]{aastex63}

\usepackage{ulem}
\usepackage{amsmath}

\shorttitle{IAGB Stars I.}
\shortauthors{Freedman et al.}
\graphicspath{{./}{figures/}}

\begin{document}
\title{I-Band Asymptotic Giant Branch (IAGB) Stars:\\ I. Exploring a New Standard Candle for the Extragalactic Distance Scale\footnote{Based on observations made with the NASA/ESA Hubble Space Telescope, obtained at the Space Telescope Science Institute, which is operated by the Association of Universities for Research in Astronomy, Inc., under NASA contract NAS 5-26555.}}

\author{Barry~F.~Madore}\affil{Observatories of the Carnegie Institution for Science 813 Santa Barbara St., Pasadena, CA~91101}\affil{Department of Astronomy \& Astrophysics, University of Chicago, 5640 South Ellis Avenue, Chicago, IL 60637}

\author{Wendy~L.~Freedman}\affil{Department of Astronomy \& Astrophysics, University of Chicago, 5640 South Ellis Avenue, Chicago, IL 60637}\affiliation{Kavli Institute for Cosmological Physics, University of Chicago,  5640 South Ellis Avenue, Chicago, IL 60637}\correspondingauthor{Wendy L. Freedman}\email{wfreedman@uchicago.edu}

\author{Taylor Hoyt}\affil{Department of Astronomy \& Astrophysics, University of Chicago, 5640 South Ellis Avenue, Chicago, IL 60637}

\author{In Sung~Jang}\affil{Department of Astronomy \& Astrophysics, University of Chicago, 5640 South Ellis Avenue, Chicago, IL 60637}\affiliation{Kavli Institute for Cosmological Physics, University of Chicago,  5640 South Ellis Avenue, Chicago, IL 60637}

\author{Abigail~J.~Lee}\affil{Department of Astronomy \& Astrophysics, University of Chicago, 5640 South Ellis Avenue, Chicago, IL 60637}\affiliation{Kavli Institute for Cosmological Physics, University of Chicago,  5640 South Ellis Avenue, Chicago, IL 60637}
\author{Kayla A. Owens}\affil{Department of Astronomy \& Astrophysics, University of Chicago, 5640 South Ellis Avenue, Chicago, IL 60637}

\begin{abstract}
In the I-band color-magnitude diagrams (CMD) of resolved nearby galaxies, the reddest asymptotic giant branch ({\it AGB}) stars form a previously unremarked-upon, but nevertheless distinct and easily-identified population of high-luminosity stars. Hereafter we refer to this population as being comprised of {\it I-Band AGB} (IAGB) stars. Identifying these stars in the Large Magellanic Cloud (LMC), the Small Magellanic Cloud (SMC) and in NGC~4258 (for all three of which there are published geometric distances) we find that the marginalized luminosity functions are each well approximated by single-peaked Gaussians, having one-sigma dispersions of +/-0.22~mag, +/-0.25~mag and +/-0.24 mag, respectively. The zero points for the modal I-band absolute magnitudes of {\it IAGB} stars are found to be $M_I =  -4.49 \pm$0.003~mag (stat) in the LMC (4204 stars), $M_I =  -4.67 \pm$0.008~mag (stat), for the SMC sample (916 stars), and $M_I =  -4.78 \pm$0.030~mag (stat) for NGC~4258 (62 stars).  A global average over these three independent calibrations of the {\it IAGB} zero point (weighted inversely by squares of their systematic errors) gives  $<M_I>~ =  -4.65 \pm$  0.119~mag (stat) +/- 0.025 (sys). In Paper II we will show the results of applying the {\it IAGB Method} to 92 galaxies additional galaxies resolved by HST, reaching out to distances just short of 10 Mpc.  
\end{abstract}

\section{Introduction}
 {\it J-Branch AGB} stars (Nikolaev \& Weinberg 2000, Weinberg \& Nikolaev 2001), defined photometrically by their near-infrared (NIR) colors, are now known to be powerful distance indicators, having mean J-band luminosities of $M_J = $ -6.2 mag (Freedman \& Madore 2020, Madore \& Freedman 2020, Ripoche et al. 2020; followed up by Parada et al. 2021, 2023; Lee et al. 2021a,b; Lee et al. 2022; Lee 2023 and references therein).
By direct analogy, we explore here the possibility that, at slightly bluer wavelengths there are {\it AGB} stars (not necessarily dominated by carbon stars, as is the case for near-infrared-color {\it JAGB} stars), that might also qualify as standard candles, suitably chosen by their (red) colors and magnitudes. 

The history of attempts to specifically isolate carbon stars in optical bands is covered in a review by Battinelli \& Demers (2005). Here our motivating rational is not premised on finding and using carbon stars {\it per se}, but rather in finding and using a similar population of bright red stars that can be uniquely defined by their (distance-independent) colors and that, when averaged over an adequate sample, can act as high-precision standard candles, as is the case for the JAGB stars.  

Here we identify such a population of stars, plainly (in retrospect) visible in the optical (VI) color-magnitude diagrams (CMDs) of nearby galaxies, as already surveyed by a number of programs using the Hubble Space Telescope (HST) and its optical imagers (WFPC-2, ACS and WFC3-UVIS). In this, the first in a series of papers on the topic, we have identified this population of stars, specifically in ground-based I vs (V-I) CMDs of the two nearby Local Group galaxies, the Large and Small Magellanic Clouds (LMC and SMC, respectively), and in HST/ACS F814W imaging of the more distant galaxy, NGC~4258. We suggest preliminary color cuts in order to quantitatively define these stars photometrically (without recourse to spectroscopy and/or narrow-band imaging). The color cuts have been chosen here to minimize the contribution of the oxygen-rich AGB population to the blue, while maximizing the contribution of the carbon-rich AGB population to the red. (To some degree this flexible choice of the color cut will respond to the additional reddening of this AGB disk population. A deeper discussion of this topic is warranted, and will be addressed in future papers.) In all three galaxies, the marginalized I-band luminosity functions are found to be single-peaked Gaussians, with a smooth, and more widely dispersed, background level of other stars (and possibly a contribution from unresolved, point-like background galaxies) acting as a pillar, but not changing the mode of the distribution. Both galaxies have independently determined ``geometric'' distances (see below), which we adopt in the process of calibrating the absolute zero point of this population of stars which we call ``I-band Asymptotic Giant Branch" (hereafter {\it IAGB}) stars.
\begin{figure*}
\centering
\includegraphics[width=1.0\textwidth]{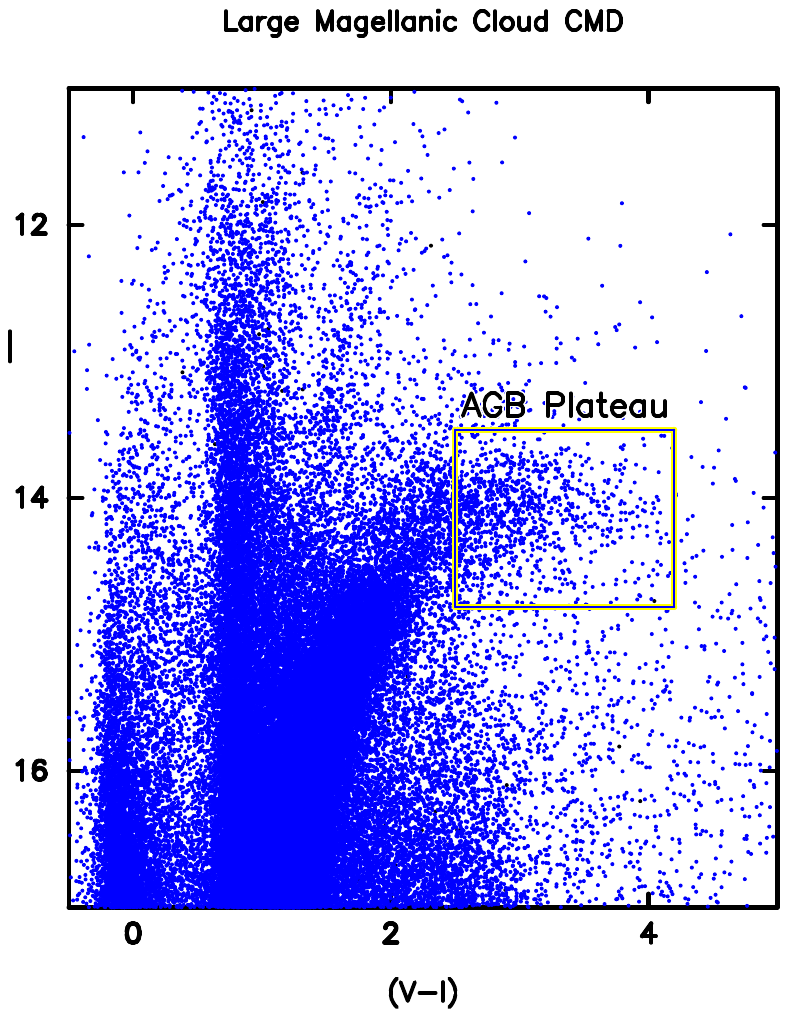}
\caption{I vs (V-I) Color-Magnitude diagram for the Large Magellanic Cloud. The newly identified AGB Plateau population of stars, discussed in this paper, is highlighted in black and yellow.  Its flat luminosity function and unique color selection makes it an ideal candidate for being an extragalactic distance indicator.} 
\label{fig:f1}
\end{figure*}

In Paper II we present our comparison of Tip of the Red Giant Branch ({\it TRGB}) and {\it IAGB}  apparent magnitudes for 92 nearby galaxies, for which both types of distance measurements are simultaneously  available in the same galaxy. There are advantages to this particular starting point:
(1) Components of the total line-of-sight extinction that are common to each of the two distance indicators (the MW foreground extinction, for instance) will simply displace points equally along a line of unit slope, leaving the mean {\it intrinsic} offset between them untouched, and the scatter in the global correlation also unaffected.  
(2) The mean apparent zero-point offset between the two methods will calibrate the mean apparent, absolute magnitude of the {\it IAGB} stars. 
Assuming that this sample of 92 calibrating galaxies is representative of target galaxies as a whole, then any additional extinction that the {\it IAGB} stars may be encountering, in excess of the {\it TRGB} line of sight, will be included in the calibrators and will cancel out (with some scatter, to be estimated) {\it in the mean}; the different star formation histories will also cancel in the mean, and the different metallicities of the {\it IAGB} populations will in turn cancel out, in the mean. All of these residual uncertainties will be found in the scatter in the differences between the {\it TRGB} and {\it IAGB} apparent magnitudes. Our expectation is that if any one of these parameters scales with its mean offset then the observed scatter will be indicative of the potential for bias somewhere in the mix. Inversely, small scatter will indicate that the calibrating sample has uniformity in these nuisance parameters and that their mean values will not introduce excessive bias when applied to new samples.  Here, our goal is much more focused, primarily on the zero point. We begin with the Large Magellanic Cloud (LMC).

\subsection{I-Band AGB Stars in the Large Magellanic Cloud}

Figure 1 shows a broadly-based view of the optical (VI) CMD of the intrinsically brightest stars in the LMC (Zaritsky et al. 2004). The reddest and apparently brightest feature in the CMD is labeled {\it AGB} Plateau. All other stellar populations are bluer than this feature, and no populations of any significance are found below it, to the magnitude limit of this CMD.

The Large Magellanic Cloud has the following attributes as a calibrator for this new method: (a) The LMC has a large number of detached eclipsing binaries (DEBs) contributing to the geometric distance to this galaxy, 20 in all (Pietrynzski et al. 2019). The DEB stars are relatively concentrated around the central bar of the LMC, plausibly giving an unbiased average distance to that region in which we also measure the {\it IAGB} stars. (b) The extinction along this particular LMC line of sight is non-negligible and has become a topic of much discussion in other contexts (see Hoyt (2023) and references therein). Here we have chosen to be consistent with our published analysis of the extinction adopted for the TRGB stars in this same region (Freedman et al. 2019). This choice remains consistent with the most recent dust maps of Scowron et al. (2022), and as well as the detailed analysis undertaken by Hoyt cited above, and is self-consistent across our publications, including different distance determination methods. (c) The LMC is relatively face-on and accordingly the DEBs, {\it TRGB} population, the {\it IAGB} stars, (selected to be along the sames lines of sight) are not expected to have significant differences in their mean distances.

\begin{figure}
\centering
\includegraphics[width=0.8\textwidth]{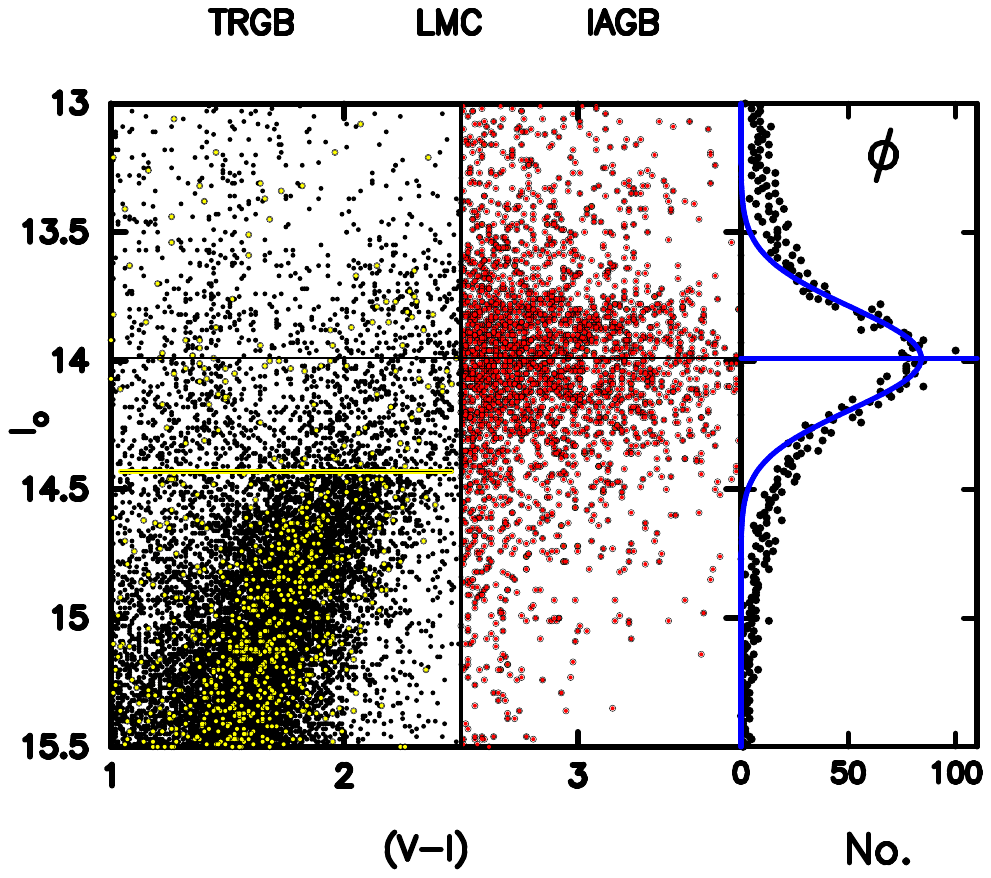}
\caption{The central region around the mode of the smoothed {\it IAGB} luminosity function for the LMC (right-most panel: large, black dots, binned in 0.01~mag intervals) is well fit by the symmetric Gaussian shown in blue, having a dispersion of +/-0.20~mag around an apparent modal value of $I$ = 14.16 mag. Given the sample size of 4,204 {\it IAGB} stars (within two sigma of the mean), the error on the modal value is found to be +/-0.003~mag. The horizontal black line crossing the two panels, shows the mode of the {\it IAGB}. The horizontal yellow line below it marks the level of the {\it TRGB} for reference and comparison. One in ten stars are plotted in yellow so as to give a better sense of the density distribution along and across the RGB. 
All stars in the CMD redder than (V-I) = 2.5~mag are shown as red dots. }
\label{fig:f1}
\end{figure}

\begin{figure}
\centering
\includegraphics[width=0.6\textwidth]{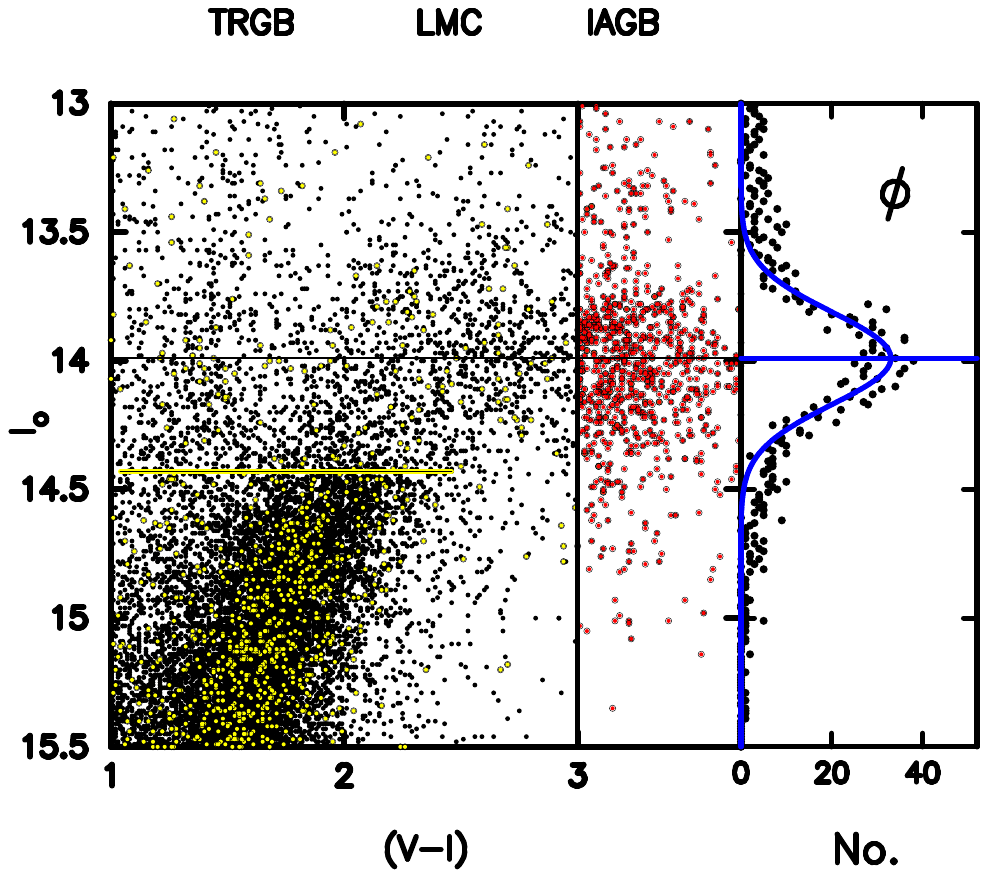}
\includegraphics[width=0.6\textwidth]{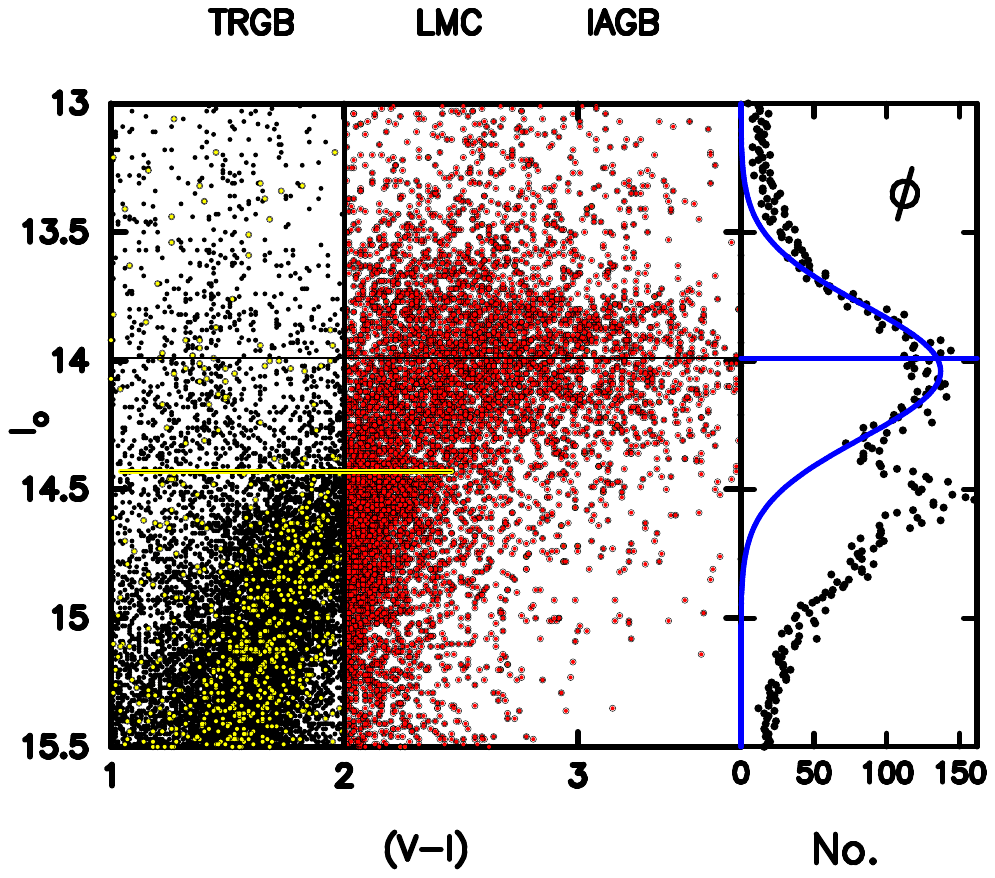}
\caption{Same as Figure 2 except that the upper panel shows the results for a color cut at (V-I) = 3.0 mag, and the lower panel shows a color cut at (V-I) = 2.0~mag. The mode of the IAGB is unchanged for the redder cut, while the mode for the bluer cut shifts fainter by only 0.04 mag.
The blue horizontal line in the right half of the lower panel shows the modal value from Figure 2. The widths of the fitted Gaussian systematically change, with the redder cut being 0.18~mag and 0.26~mag for the bluer cut.}
\label{fig:f2}
\end{figure}

In Figures 2 and 3 we zoom in on a region in the CMD that is immediately around the {\it IAGB} population, explicitly imposing, in Figure 2, a preliminary blue cut-off at (V-I) = 2.5~mag, and marginalizing (in magnitude) all of the stars redder than this color. The latter subset of stars is shown as an I-band luminosity function, plotted in the panel to the right of the CMD. A Gaussian with a sigma of $\pm0.22$ mag, centered on the mode of the {\it IAGB} luminosity function, was fit to the core of the unbinned data shown in the right panel of Figure 2. These data were dereddened  a value of $A_I = $ 0.16~mag as determined independently by Hoyt (2023)\footnote{For uncertainties in the extinctions for each of the anchor galaxies we have adopted Hoyt's values +/-0.017 \&  0.022 mag for the LMC and SMC, resp. For NGC 4258 we simply adopted a comparable uncertainty of +/-0.020~mag.} The mode of the {\it IAGB} luminosity function, marked by the horizontal blue line, is at $I_o = $ 13.99~mag +/- 0.003~mag (standard error on the mean), given the standard error on the distribution (+/-0.22~mag) and the 4,042 stars within +/-2 sigma of the mode. The stability of the mode and variations in the dispersion in response to large (+/-0.5~mag) variations in the blue cutoff in IAGB luminosity function are shown in the two panels of Figure 3.

Using the DEB geometric distance modulus of 18.477~  +/-~0.024 mag (sys) to the LMC (Pietrzy\'nski et al. 2019)  this gives us the first (and most statistically significant) zero-point calibration of the
{\it IAGB} with an absolute magnitude
of $$M_I(LMC) =  14.15 - 0.16 - 18.477$$
$$= -4.487~ +/-0.003~ (se)~+/-0.024~ (sys)~mag.$$

\subsection{I-Band AGB Stars in the Small Magellanic Cloud}

The pros and cons of the SMC are almost orthogonal to those of the LMC: (a) The number of DEBs known and measured in the SMC (Graczyk et al. 2014) is significantly fewer (but non-negligible) in comparison to the LMC (Pietrzy{\'n}ski et al. 2019) sample (20 versus 4). (b) The total line-of-sight extinction is generally thought to be factors smaller than the extinction towards the LMC distance indicators being considered here (Hoyt 2023). And finally, (c) The SMC distance tracers may all be influenced by the measurable back-to-front geometry of this tidally disturbed galaxy (e.g., Scowcroft et al. 2016 and references therein).

Figure 4, left panel, shows the I vs (V-I) CMD for stars in the SMC taken from Zaritsky et al. (2002). The  vertical black line at (V-I) = 2.0~mag marks the color beyond which stars are chosen for the marginalized luminosity function, which is shown by the unsmoothed black line shown in the right panel. Additional details are given in the caption.  There are  916 {\it IAGB} stars found within two sigma (i.e., +/- 0.44~mag) of the modal magnitude, I = 14.37~mag. This gives a standard error on the mean of 0.22/$\sqrt(915)$ = 0.007~mag. Correcting for a foreground I-band extinction of $A_I = $0.06~mag (NED Foreground Extinction Calculator 2024, based on Schlafly \& Finkbiner 2011 and Schlegel et al. 1998) and subtracting a true distance modulus of 18.977 +/-~0.028~mag (sys), based on 4 geometric distances to detached eclipsing binaries in the SMC  (Graczyk et al. 2014) results in our second estimate for the zero point of the {\it IAGB} method, of  $$M_I (SMC) = 14.37 - 0.06 - 18.977$$
$$=  -4.667~ +/- 0.007~(se)~ +/- 0.028~ (sys)~mag.$$

\begin{figure*}
\includegraphics[width=0.7
\textwidth]{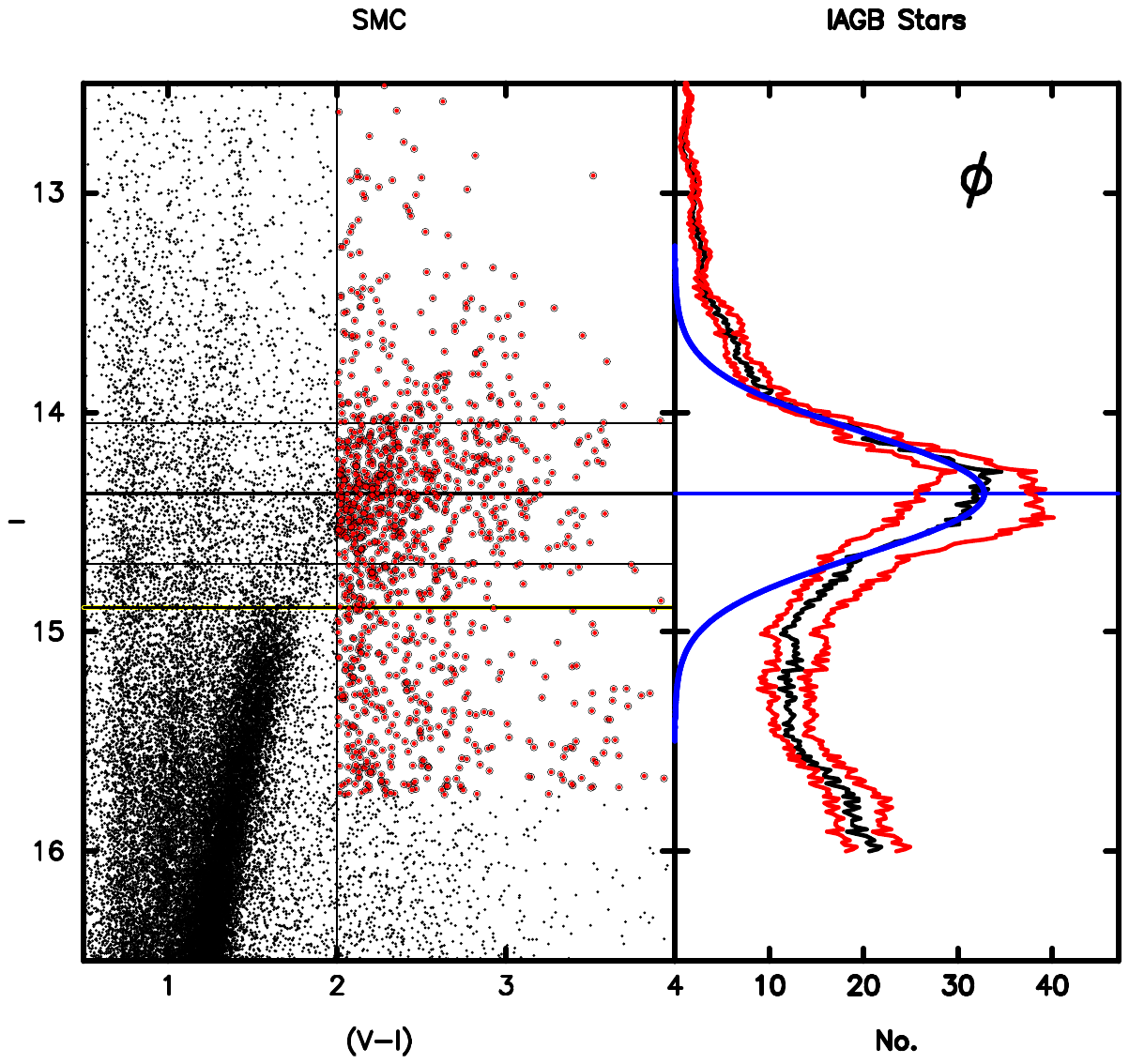}
\label{fig:f3}
\caption{Selection of {\it IAGB} stars in the Small Magellanic Cloud. The I-band CMD for stars in the SMC is shown in the left panel. The marginalized luminosity function of the stars with (V-I) $>$ 2.0~mag (black line) is shown in the right panel. The red luminosity functions are the result of  changing the red color cutoff by +/-0.1~mag so as to give a visual indication of the sensitivity of the modal value to the adopted blue cutoff. The blue line is a Gaussian fit to the core of the black luminosity function having a sigma of 0.22~mag and a maximum occurring at an apparent magnitude of I = 14.37 +/- 0.22~mag. 916 IAGB stars are found within two sigma of the modal magnitude, resulting in an error on the mean of 0.25/$\sqrt(915)$ = 0.008~mag. The dashed blue line indicates the level of the TRGB discontinuity, emphasizing the brighter 0.5 mag advantage of the {\it IAGB} over the {\it TRGB} as a distance indicator.} 
\end{figure*}

\begin{figure*}
\includegraphics[width=0.8\textwidth]{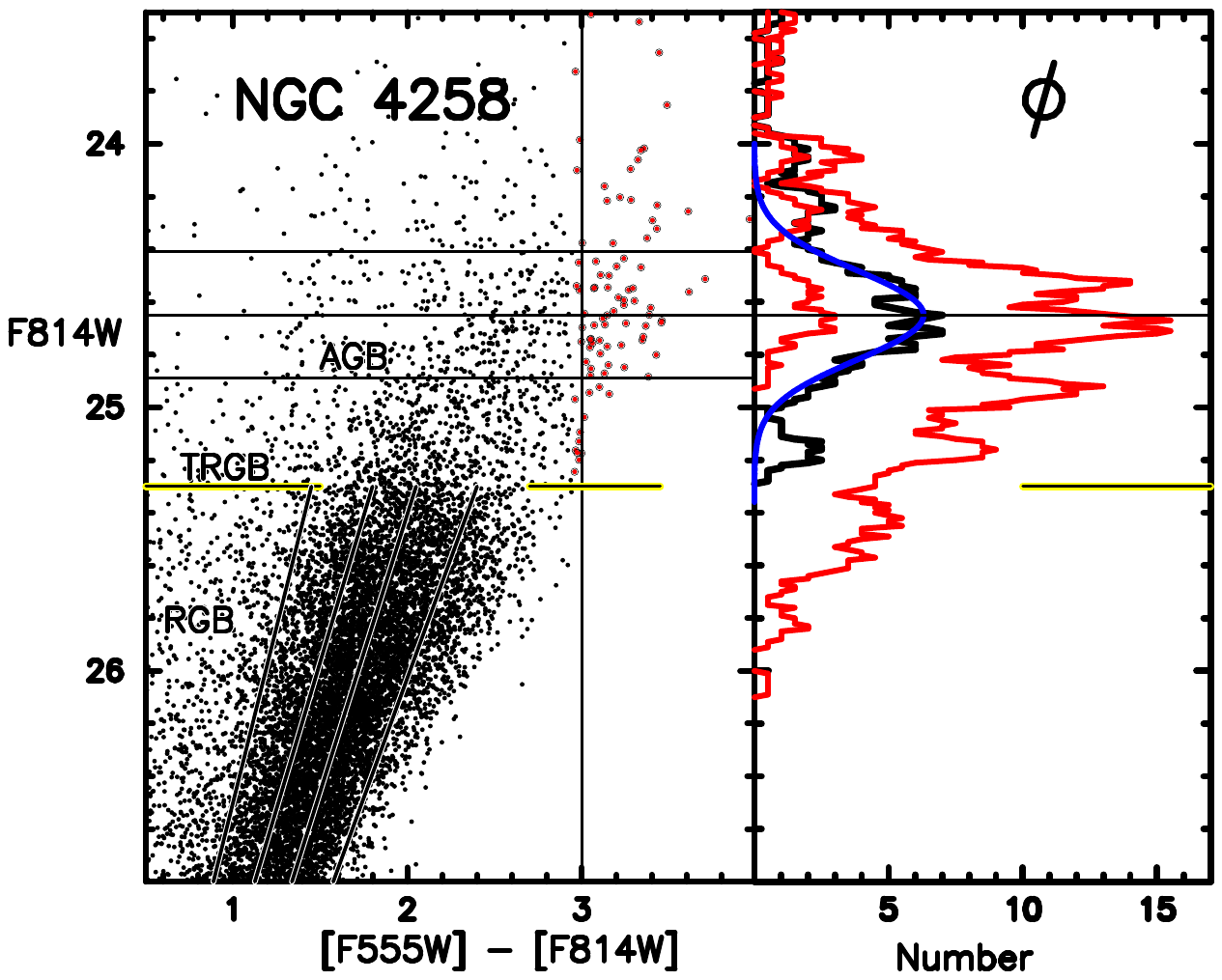}
\caption{Detection and measurement of the {\it IAGB} population in NGC~4258 as marked by the circled red points in the left panel. Marginalized luminosity functions for the color-selected {\it IAGB} stars are shown in the right panel. The black line is the adopted curve fit by a Gaussian (the thick black line) centered at I = 24.65~mag, having a sigma of 0.24~mag. The red lines to either side of the adopted luminosity function are the result of shifting the blue cut-off by +/-0.1 mag, so as to illustrate the effect of that choice on the adopted modal value of the {\it IAGB} luminosity.}
\label{fig:f4}
\end{figure*}

\begin{figure*} \centering
\includegraphics[width=0.8\textwidth]{"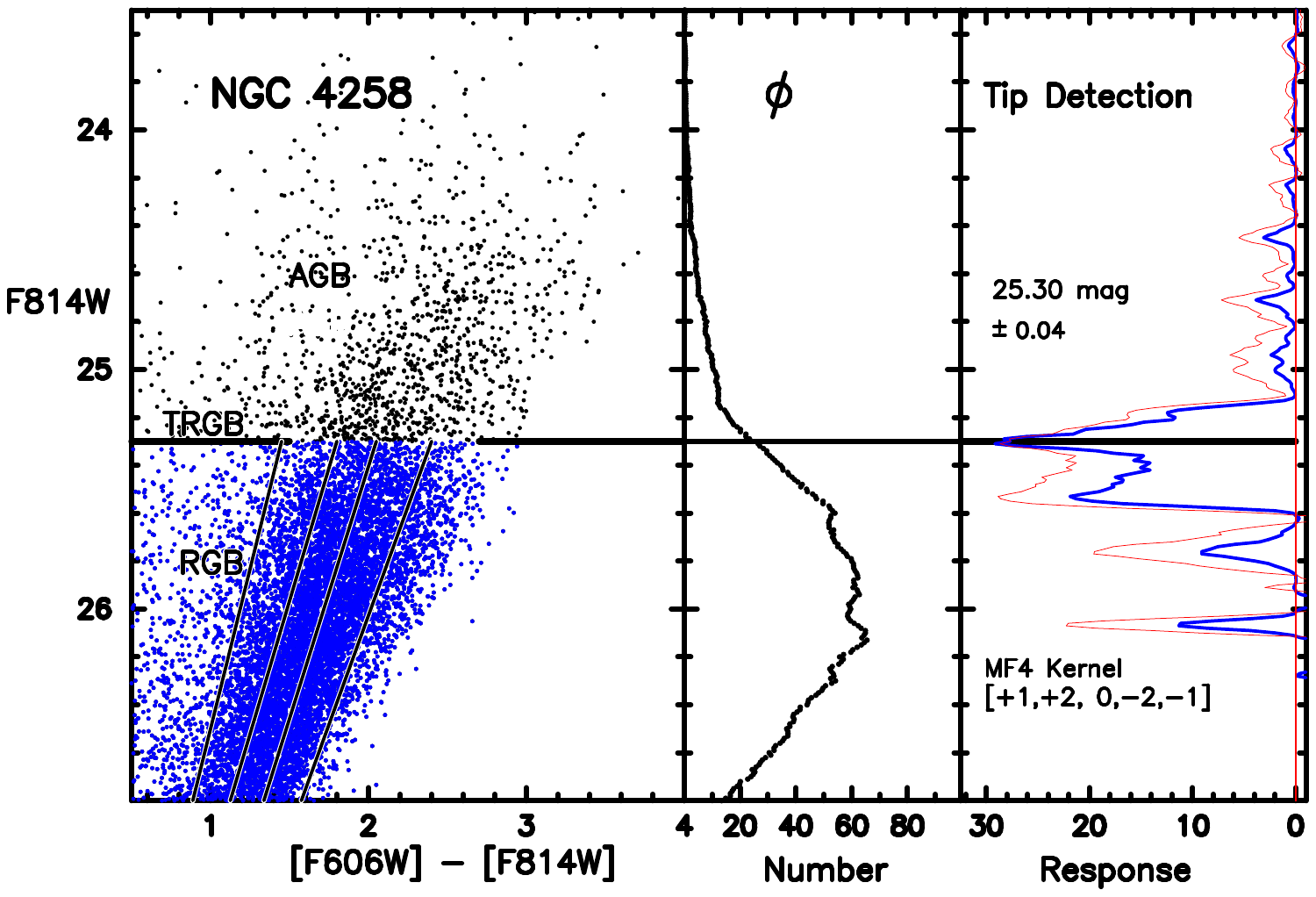"}
\caption{Detection and measurement of the apparent I-band magnitude of the {\it TRGB} in NGC~4258. Left Panel: The CMD showing {\it RGB} stars in blue, The {\it TRGB} is marked by the two black horizontal lines at I = 25.30~mag crossing the entire figure. Middle Panel: The smoothed luminosity function of the stars in the CMD to the left. Right Panel: The edge-detection response (blue line) to passing an MF4 kernel (a modified Sobel filter, taken from Madore et al. 2023). The peak response  marks the TRGB. The fainter red line is the unweighted fit; the blue line is the noise-weighted response. Note that the noise-weighted response is unbiased for all fluctuations; but, it noticeably suppresses false-positives, which are due to Poisson fluctuations in the number counts, and enhances the contrast only for the true {\it TRGB}, as previously demonstrated in the extensive simulations of Madore et al. (2023). 
}
\label{fig:f5}
\end{figure*}

\subsection{I-Band AGB Stars in NGC~4258}
The mega-maser galaxy, NGC~4258, is our third and final geometric, zero-point calibrator for the {\it IAGB} method. In comparison to the Magellanic Clouds, the much more distant galaxy, NGC~4258 is unique in many respects: (a) NGC~4258 has only one geometric distance determination to it (Reid et al. 2019), but that one application shares no systematics with the DEB method used in the other two calibrating galaxies. NGC~4258 remains the only maser galaxy close enough to have {\it TRGB}, Cepheid, {\it JAGB} and/or  {\it IAGB} apparent magnitudes measured. (b) Reddening corrections for the {\it TRGB} stars in NGC~4258 are not a problem, if one obeys the stricture of observing those stars only in the dust-free halo (as we have done in our publications; nevertheless, others disagree with this course of action, e.g., Wu et al. 2023). 
(c) NGC~4258 is sufficiently distant (7 Mpc, or about 140 times further away than the Magellanic Clouds) that there is no issue with its 3-dimensional geometry.
(d) Globally, the metallicities of the {\it IAGB} stars in NGC~4258 are expected to be higher than either their LMC or SMC counterparts, but no direct determinations for these specific types of stars have been made in the outer-disk field used here. { However, if the metallicity of this given field is particularly low, that could result in the depleted population of carbon stars seen here given the known, positive correlation of the C/O ratio (Carbon-rich to Oxygen-rich) with metallicity (see Hamren et al. 2015 for a discussion and a recent application). More densely Carbon-star populated regions in NGC~4258 will be sought out in future observations.} 

The CMD for NGC~4258, and its marginalized I-band luminosity function are shown in Figure 5. Blue points show the selected {\it IAGB} stars and their luminosity function is shown by the blue line in the left panel. Approximately 0.5 magnitudes fainter than the modal value of the {\it IAGB} is the {\it TRGB}, marked by black-on-yellow horizontal lines, in the two panels.
The TRGB and its tip detection are further highlighted, for reference, in Figure 6. 

The apparent I-band magnitude of the {\it IAGB} is I = 24.65~mag and the best-fitting Gaussian to the luminosity function has a sigma of +/-0.24~mag defined by 62 {\it IAGB} stars within 2 sigma of the mode; leading to a sigma on the mean of 0.24/$\sqrt(61)$ = +/-0.031~mag. Correcting the apparent magnitude for a foreground extinction of $A_I = $ 0.03~mag (NED Foreground Extinction Calculator 2023, based on Schlafly \& Finkbiner 2011 and Schlegel et al. 1998) and subtracting the true geometric distance modulus of 29.397~mag (Reid et al. 2019) gives us our third estimate of the zero point of the {\it IAGB}  distance scale. That is $$M_I(N4258) = 24.65 - 0.03 - 29.397$$
$$= -4.777~+/- 0.031~ (se)~ +/- 0.022~ (sys)~mag.$$
\vfill\eject
\subsection{The Averaged Zero Point for the IAGB Method}

Table 1 summarizes the three independent estimates given above for the zero point of the {\it IAGB} population of stars. The result is 
$$M_I (IAGB) = (-4.487 -4.667 -4.777)/3$$
$$= -4.644 +/- 0.079~(se)~ +/-0.028~(sys)~mag,$$

where the statistical error (0.116 mag) is the unweighted average of the scatter of the three absolute magnitudes about their mean. The standard error on the mean is then +/-0.116/SQRT(N-1) = 0.079, where N = 3. The systematic error on the mean is the quadrature-summed  average of the individual systematic errors of each of the three zero-point calibrators.

\clearpage
\section{Discussion and Summary}\label{sec:sec6}

\begin{deluxetable*}{crccclcccc}
\tablecaption{IAGB Zero Point Calibrators}\label{tab:observ}
\tablehead{
\colhead{Name} & {No.} & {I} &
{$\sigma$(sem)} 
& {$A_I$} &
{$\sigma_{A_I}$} &  {mod$_o$} &{$M_I$} &{$\sigma$(stat)} &{$\sigma$(sys)}\\
&{}&{mag}&{mag}&{mag}&{mag}&{mag}&{mag}&{mag}&{mag}}
\decimals
\startdata
LMC &  4204 & 14.15  & 0.003 & 0.16 & 0.017 &18.48 & -4.49 & 0.017 & 0.024\\
SMC &  916  & 14.37  & 0.007 & 0.06 & 0.022 &18.98 & -4.67 & 0.023 & 0.028\\
NGC~4258 &  63 & 24.65 & 0.030 & 0.03 & 0.020 &29.39 & -4.77 & 0.036 &0.022 \\
\enddata
\end{deluxetable*}

Based on three independent estimates of the zero-point of the {\it IAGB} stars in the LMC, SMC and NGC~4258, we determine an average I-band absolute magnitude of $M_I =  -$4.64 +/- 0.116 (stat) +/- 0.082 (sem) +/- 0.028 (sys), where the statistical error in Column 9 is the quadrature sum of the the error on the apparent magnitude of the tip and the uncertainty on the extinction.

There is significant statistical scatter (+/-0.116~mag) between the three zero-point calibrators, when compared to the quoted individual statistical errors given in Table 1, that still needs to be understood.  There are  many conceivable factors that could be contributing to this, including internal reddening differences between the three galaxies, differences due to metallicity and differences due to variation from galaxy to galaxy in their star formation histories. A simple rank ordering of the galaxies by metallicity has NGC~4258 being the most metal rich, followed by the LMC and finally the SMC at the most metal poor end. The zero points, however, are rank ordered NGC~4258, SMC and then LMC, brightest to faintest. Thus no metallicity effect is immediately apparent, but a larger sample is needed to test for a significant effect. Both metallicity and star formation history will be explored in Paper II, where we put a more robust upper limit on the impact of these, and any other potential systematic effects.

It is worth noting, however, that the scatter in the {\it IAGB}, quoted here, is comparable to, or slightly smaller than, the intrinsic scatter in the Type~Ia SNe calibration, which is at the level of +/-0.13-0.18~mag (Burns et al. 2019).

\acknowledgments
We thank the {\it Observatories of the Carnegie Institution for Science} and the {\it University of Chicago} for their support of our long-term research into the calibration and determination of the expansion rate of the Universe.

\end{document}